\documentclass[journal,9pt]{IEEEtran}
\usepackage[utf8]{inputenc}
\usepackage{amsmath,amsfonts,amssymb,empheq,mathtools,bm}
\usepackage{algorithmic}
\usepackage{algorithm}
\usepackage{array}
\usepackage[caption=false,font=footnotesize]{subfig}
\usepackage{textcomp}
\usepackage{stfloats}
\usepackage{url}
\usepackage{verbatim}
\usepackage{graphicx}
\usepackage{xcolor}
\usepackage{cite}
\usepackage{booktabs}
\usepackage{multirow}
\usepackage{makecell}
\usepackage[breaklinks,hidelinks]{hyperref}
\usepackage{lipsum}
\hypersetup{colorlinks=true}
\usepackage[breaklinks]{hyperref}
\usepackage[normalem]{ulem}

\definecolor{redX}{HTML}{D40200}
\definecolor{violetX}{HTML}{9D1569}
\newcommand{\dd}{\,\mathrm{d}}

\newcommand{\taumin}{\tau_{\min}}

\begin{document}

\title{\huge{Spectrally smooth broadband response via autocorrelation-constrained inverse design}}

\author{\Large{Johannes Gedeon, Rasmus E. Christiansen, and Ole Sigmund}
\thanks{
This work was supported by VILLUM FONDEN through the Villum Investigator Project AMSTRAD (VIL54487).}
\thanks{J. Gedeon, R. E. Christiansen, and O. Sigmund are with the Department of Civil and Mechanical Engineering, Technical University of Denmark, Nils Koppels All{\'e} 404, DK-2800 Kgs. Lyngby, Denmark (e-mail: joged@dtu.dk; raelch@dtu.dk; olsi@dtu.dk).}%
\thanks{R. E. Christiansen and O. Sigmund are with NanoPhoton -- Center for Nanophotonics, Technical University of Denmark, {\O}rsteds Plads 345A, DK-2800 Kgs. Lyngby, Denmark.}}

\markboth{IEEE Transactions on Antennas and Propagation}%
{Gedeon \MakeLowercase{\textit{et al.}}: Spectrally Smooth Broadband Response via Autocorrelation-Constrained Inverse Design}

\maketitle

\begin{abstract}
Time-domain inverse design in photonics is known to be suitable for maximizing the efficiency of optical devices over broad frequency ranges. Objectives commonly used in this context include time-integrated field quantities derived from Poynting's theorem, such as energy, flux, or dissipated power, which can be directly linked to integrated frequency-domain responses via Parseval's theorem. While computationally efficient, these objectives measure only the total response over the targeted bandwidth and, as we demonstrate, are insufficient to capture undesired in-band ripple, narrow spectral features, or sidelobes.  We overcome this limitation by introducing a time-domain metric quantifying such spectral variations based on the weighted long-lag autocorrelation energy of the optical response. We incorporate this metric into an FDTD-based topology-optimization framework and demonstrate its beneficial effect on the example of inverse designing one-dimensional dielectric Bragg mirrors via the time-domain adjoint method.

\end{abstract}


\section{Introduction}\label{Sec:Introduction}

\IEEEPARstart{O}{ptical} devices designed to operate across broad frequency ranges are important for a variety of applications, including solar energy harvesting, broadband absorbers, dielectric mirrors, and achromatic imaging~\cite{Mascaretti2023,Yu2019BroadbandAbsorbers,Chang2024BroadbandMirrors,Chen2018AchromaticMetalens}. Therefore, there is strong interest in developing advanced inverse design techniques to maximize broadband performance. The development of gradient-based optimization algorithms that rely on Maxwell's equations for electromagnetics in the time domain has proven efficient for tackling the broadband response, as the time-domain gradient captures the sensitivities across the full range of frequencies contained in a tailored excitation pulse~\cite{AdaptivePulse}. One possible approach to define a time-domain objective is to prescribe an explicit target signal and minimize the difference between the simulated response and the reference, integrated over time~\cite{PulseShaping}. While this can provide precise control over the temporal response, it requires prior knowledge of the desired waveform and can make the inverse-design problem restrictive and challenging. A simpler approach is to optimize time-integrated quantities of the simulated response field directly, such as the electric-energy density \(\frac{1}{2}\varepsilon|\mathbf E|^2\) or the dissipated-power density \(\sigma|\mathbf E|^2\)~\cite{AdaptivePulse,NomuraAntennas,EmadDissipation,GedeonPowerDissip}, which are natural objectives for broadband field confinement, reflection, or absorption. By Parseval's theorem~\cite{Bracewell2000Fourier}, such quadratic time-integrated quantities are directly related to corresponding frequency-integrated field responses. The main trade-off is the loss of phase information and the limited control over how the response is distributed across the target bandwidth, which can lead to undesired spectral variations. To address this, Park et al.~\cite{AdaptivePulse} recently proposed an adaptive spectral-weighting strategy based on dynamically tuning the incident pulse and demonstrated improved spectral uniformity. However, it requires explicit identification of underperforming spectral regions and modification of the excitation pulse during the iterative optimization process.

We propose an alternative approach to tackle this problem without the need to adapt the excitation pulse, specifically aimed at suppressing sharp and localized spectral features. It relies on a time-domain metric that measures the weighted long-lag tail energy of a signal's autocorrelation function and is related to the weighted integrated sidelobe level~(WISL) used in signal processing~\cite{WISL}. Since the autocorrelation and the energy spectral density form a Fourier-transform pair~\cite{Bracewell2000Fourier}, the metric provides an additional time-domain measure of these spectral features over a broad band of frequencies. It thereby addresses a central limitation of integrated time-domain objectives by providing direct sensitivity to localized spectral defects that may otherwise remain effectively invisible to the optimizer. We demonstrate this capability on the example of maximizing broadband reflectance of 1D~dielectric gratings, incorporating the metric as either a constraint or a penalization term.

\begin{figure}[!b]
    \centering
    \includegraphics[trim=0mm 0mm 0mm 0mm, width=\columnwidth]{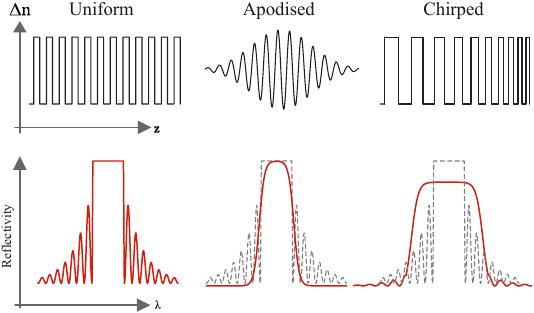}
    \caption{Schematic illustration of different index modulations in Bragg gratings and their resulting reflectance spectra. The figure is adapted from Ref.~\cite{Ams-gratings}.}
    \label{Fig:BraggGratings}
\end{figure}

Bragg gratings are among the most well-studied structures in the photonics literature and are predominantly used in fiber-optic engineering~\cite{Othonos1997Review}. Different non-trivial designs have been proposed to achieve efficiency across a wide range of frequencies (chirped gratings) or suppress undesirable sidelobes in the reflection response (apodized gratings), cf. Fig.\,\ref{Fig:BraggGratings}. Therefore, they serve as an ideal test case for studying the effect of our metric in optimizations targeting broadband efficiency and a smooth broadband response. As the inverse design method, we choose density-based topology optimization (TopOpt)~\cite{RasmusTutorial}, which allows us to explore the full design space and allows us to converge to structures exhibiting a continuous modulation of the refractive index. As a time-domain solver, we choose the Finite-difference time-domain (FDTD) method, which is a popular numerical method for solving Maxwell's equations in time and also allows evaluating the time-harmonic responses over a desired bandwidth~\cite{Taflove}. 

The code underlying the results presented in the following is openly available as a GitHub repository~\cite{Grating_TimeOpt_Github}.

\section{A time-domain metric for spectral smoothness}\label{Sec:TAC}

\begin{figure}[!t]
    \centering
    \includegraphics[trim=0mm 0mm 0mm 0mm, width=\columnwidth]{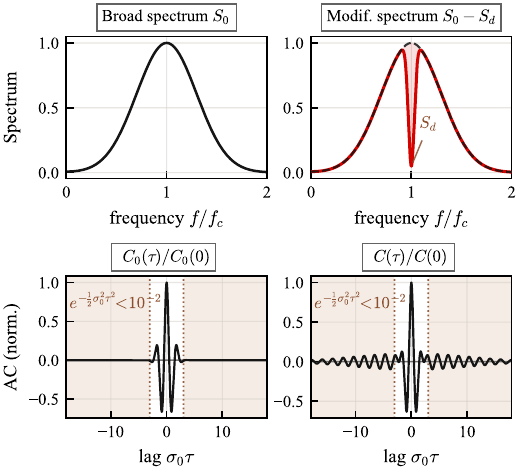}
    \caption{Illustration of the Fourier correspondence between a narrow spectral feature in a broadband spectrum and a long-delay oscillatory tail in its autocorrelation (AC) signal, modeled by Gaussian functions. The top panels show a broad reference spectrum \(S_0\) and a modified spectrum \(S_0-S_d\), with spectral widths \(\sigma_0\gg\sigma_d\); the bottom panels show the corresponding normalized autocorrelations. The shaded region marks delays $|\tau|>\tau_{\mathrm{min}}$ for which the broad Gaussian autocorrelation envelope has decayed to less than~$1\%$ of its amplitude.}
    \label{Fig:GaussianModel}
\end{figure}

Consider a finite, real-valued continuous signal \(s(t)\) with Fourier transform~\(\hat{s}(f)\). By the Parseval--Plancherel theorem~\cite{Bracewell2000Fourier}, the signal energy can be computed equivalently in time and frequency,
\begin{align}
\int_{-\infty}^{\infty}|s(t)|^2\,\dd t
&=
\int_{-\infty}^{\infty}|\hat{s}(f)|^2\,\dd f .
\label{eq:parseval-energy}
\end{align}
This makes it attractive as a time-domain objective for broadband efficiency. However, it only measures the total spectral energy and does not constrain how this energy is distributed as a function of frequency. In particular, narrow spectral minima (e.g., of \(\pi\)-shifted gratings~\cite{piShifted}) may contribute little to the integrated energy while still compromising the $|\hat{s}(f)|$-profile over broadband. We can identify such spectral features in the time domain using the Fourier transform relation between the autocorrelation function and the energy spectral density, known as the Wiener–Khinchin theorem~\cite{Wiener1930Generalized, Bracewell2000Fourier}. Defining the spectrum as \(S(f)=|\hat{s}(f)|^2\), the autocorrelation
\begin{equation} 
C(\tau)=\int_{-\infty}^{\infty}s(t)s(t+\tau)\,\dd t, \qquad c(\tau)=\frac{C(\tau)}{C(0)} \label{eq:autocorrelation} 
\end{equation} 
forms a Fourier pair with \(S(f)\). Thus, localized spectral features correspond to slowly decaying oscillatory contributions in the autocorrelation. This relationship is illustrated schematically in Fig.\,\ref{Fig:GaussianModel}.
We start from a broad reference spectrum~\(S_0\), which gives a rapidly decaying autocorrelation \(C_0\) and is therefore characterized by a short coherence time. We then introduce a narrow spectral minimum by subtracting a localized component \(S_d\), resulting in the modified spectrum~$S_0-S_d$.
This spectral feature produces a long-lasting oscillatory tail in the autocorrelation function as a consequence.

To isolate this tail contribution, we define a reference delay \(\taumin\) from the broad spectrum \(S_0\) as the first lag at which the envelope of the normalized autocorrelation has decayed below \(1\%\) of its peak value. Delays \(|\tau|>\taumin\) therefore lie outside the central autocorrelation lobe of the broadband reference. Given a time signal \(s_0(t)\) with spectrum \(S_0\), we can compare the weighted long-lag autocorrelation energy of another signal \(s(t)\) to this broadband reference as
\begin{align}
\mathcal{T}_{\mathrm{AC}}[s;s_0]
&:=
\frac{\mathcal{E}_{\mathrm{AC}}[s]}
     {\mathcal{E}_{\mathrm{AC}}[s_0]},
\label{eq:TAC}
\\*
\text{with}\;\;
\mathcal{E}_{\mathrm{AC}}[s]
&:=
2\int_{\taumin}^{\infty}
\left(\frac{\tau}{\taumin}\right)^2
\left|c_s(\tau)\right|^2\dd\tau .
\nonumber
\end{align}
The prefactor ``\(2\)'' accounts for the symmetry~$c_s(\tau)=c_s(-\tau)$. The quadratic lag weight \(\propto \tau^2\) increases the penalty on late-time autocorrelation tails and is used to suppress delayed ringing. Since $\partial_f S(f)$ and $-i2\pi\tau C_s(\tau)$ form a Fourier-transform pair, Parseval's theorem gives \(\int |\partial_f S(f)|^2\dd f \propto \int \tau^2 |C_s(\tau)|^2\dd\tau\). Thus, \(\mathcal{E}_{\mathrm{AC}}\) can be interpreted as a time-domain measure sensitive to sharp spectral features and spectral ripples. 

For a signal \(s(t)\) with spectrum \(S_0-S_d\), as in Fig.\,\ref{Fig:GaussianModel}, the spectral feature increases \(\mathcal{E}_{\mathrm{AC}}[s]\) relative to the broadband reference and thus gives \(\mathcal{T}_{\mathrm{AC}}>1\). The condition \(\mathcal{T}_{\mathrm{AC}}=1\) includes, but is not limited to, the case in which \(s(t)\) and \(s_0(t)\) have identical spectra. We use the derived metric \(\mathcal{T}_{\mathrm{AC}}\) below for penalizing dips in the reflectance spectra using time-domain inverse design.

\section{Optimization Results}\label{Sec:Results}
\begin{figure}[h]
    \centering
    \includegraphics[trim=0mm 0mm 0mm 0mm,width=\columnwidth]{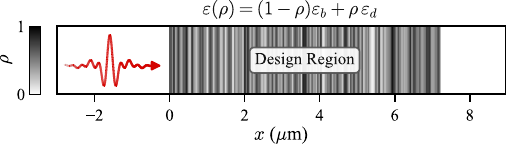}
    \caption{Setup of the optimization problem. The short pulse marked in
    red represents the broadband source, while the design region represents the
    Bragg grating and consists of the design field $\rho(x)$. The simulation domain is terminated at both ends by absorbing boundary conditions to model an unbounded domain.}
    \label{Fig:Setup}
\end{figure}

We include $\mathcal{T}_{\mathrm{AC}}$ in the inverse design of a dielectric grating for broadband reflectance. By choosing a sufficiently large material contrast and grating length, Bragg gratings can achieve nearly unity reflectance over a desired bandwidth~~\cite{Othonos1997Review}. 
We therefore choose a configuration of relatively weak gratings that makes the optimization challenging, and the broadband response remains sensitive to the detailed grating profile. The setup is shown in Fig.\,\ref{Fig:Setup}. The design region has length~\(L=7.2\lambda_0\) (\(7.2~\mu\mathrm{m}\)),
where \(\lambda_0=c/f_0=1~\mu\mathrm{m}\) is the free-space wavelength
at the center frequency \(f_0=300~\mathrm{THz}\), and is parameterized by
a density \(\rho(x)\in[0,1]\), which represents the design field in the
topology optimization. This density is mapped linearly to the grating permittivity, as indicated in Fig.\,\ref{Fig:Setup}, where~\(\rho=1\) corresponds to the presence of the design material with~\(\varepsilon_{\mathrm d}\), and \(\rho=0\) corresponds to the background medium with~\(\varepsilon_{\mathrm b}\).
Throughout this study, we use~\(\varepsilon_{\mathrm d}=1.5\) and \(\varepsilon_{\mathrm b}=1\), corresponding to a refractive-index contrast of~\(\Delta n \approx 0.22\).
\newline We inject a propagating pulse \(E_{\mathrm{inc}}(t)\) toward the grating and record the back-reflected field \(E_{\mathrm r}(t)\) from the structure as the response we use to define the figure of merit to be optimized. The pulse is a Hann-windowed sinc function modulated by a cosine carrier at the center frequency \(f_0\), such that its spectrum approximates a rectangular band of width \(2\Delta f=105~\mathrm{THz}\) centered at \(f_0\). This imposes a spectral weighting in the optimization problem across the frequencies and models a near-unity spectrum such that the energy fraction
\begin{equation}
\eta_{\mathrm r}
:=
\frac{\int E_{\mathrm r}^2(t)\,\dd t}
     {\int E_{\mathrm{inc}}^2(t)\,\dd t}
\approx
\frac{1}{2\Delta f}
\int_{f_0-\Delta f}^{f_0+\Delta f}R(f)\,\dd f ,
\label{eq:reflection-efficiency}
\end{equation}
approximates the reflectance averaged over the target bandwidth. We employ topology optimization to search for an ideal distribution $\rho(x)$ that maximizes the expression in Eq.\,\eqref{eq:reflection-efficiency}. Since $R(f)=|r(f)|^{2}$, where $r(f)$ is the complex reflection coefficient, it does \textit{not} require phase preservation across the frequencies with respect to the incident pulse.

For the post-evaluation, showing the broadband behaviour, we illustrate the reflection $R(f)$ over the target bandwidth. To quantify whether the optimized design converged to a binary grating or tended to have a continuous variation of material index, we measure the ``non-discreteness'' of the final design using~\cite{OleFilter}
\begin{equation}
   M_{\mathrm{nd}}=\frac{4}{L}\int_{0}^{L}\rho(x)\left(1-\rho(x)\right)\dd x\times 100\%,
\label{eq:M-nd}
\end{equation}
which is $100\%$ if $\rho(x) = 0.5,\ \forall x$, and 0\% if the density only consists of the values $0$ or $1$ (binary). 

For details on the gradient-based optimization employed to iteratively update~\(\rho\), as well as the TopOpt and FDTD settings used in this work, we refer the reader to Appendix~\ref{app:TopOpt}.


\subsection{Autocorrelation metric as a constraint}
\begin{figure}[!t]
    \centering

    \includegraphics[width=\columnwidth]{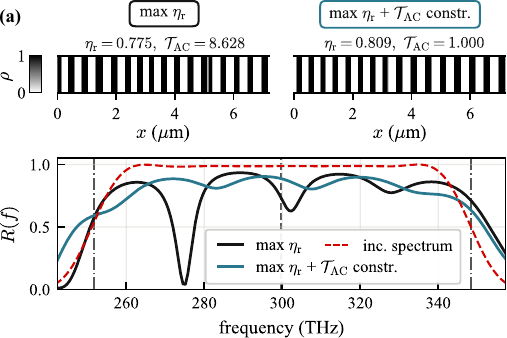}

    \par\vspace{1em}

    \includegraphics[width=\columnwidth]{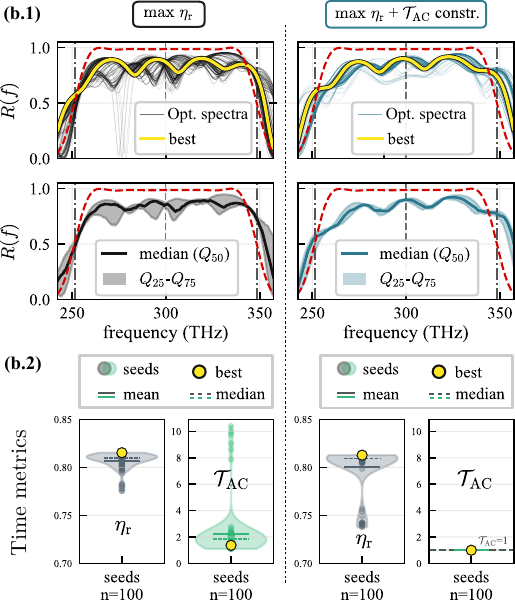}

    \caption{Optimization results comparing the unconstrained
    \(\mathcal{P}_{\mathrm{uc}}\) and constrained
    \(\mathcal{P}_{\mathrm{c}}\) formulations from Eqs.\,\eqref{eq:topopt-problems}.
    (a)~Two representative optimized gratings and reflectance spectra;
    the dashed red line denotes the incident spectrum and the vertical
    gray lines mark \(f_0\) and the half-maximum bounds defining the target bandwidth.
    (b)~Statistics over 100 topology-optimization runs initialized with
different densities:
    (b.1)~optimized reflectance spectra, with the best-performing design
    highlighted in yellow, and the pointwise median \(Q_{50}\) and
    interquartile range \(Q_{25}\)--\(Q_{75}\);
    (b.2)~distributions of \(\eta_{\mathrm r}\) and
    \(\mathcal{T}_{\mathrm{AC}}\).}
    \label{Fig:Statistics}
\end{figure}

To study the effect using the autocorrelation energy metric from Eq.\,\eqref{eq:TAC} imposed as a constraint to reduce long-tail ringing and thus smooth the reflectance response, we compare two inverse design problems:

{
\setlength{\abovedisplayskip}{1pt}
\setlength{\abovedisplayshortskip}{1pt}

\begin{subequations}
\label{eq:topopt-problems}
\begin{alignat}{4}
\mathcal{P}_{\mathrm{uc}}:\quad
& \max_{\rho}\; && \eta_{\mathrm r}
\label{eq:topopt-unconstrained}
\\
\mathcal{P}_{\mathrm{c}}:\quad
& \max_{\rho}\;&& \eta_{\mathrm r}
\notag
\\
& \phantom{1}\mathrm{s.t.} &&
\mathcal{T}_{\mathrm{AC}}[E_{\mathrm r};E_{\mathrm{inc}}]\leq 1 .
\label{eq:topopt-constrained}
\end{alignat}
\end{subequations}
}
Here, $\mathcal{P}_{\mathrm{uc}}$ denotes the unconstrained
maximization problem of the energy fraction $\eta_{\mathrm r}$ from Eq.\,\eqref{eq:reflection-efficiency}, whereas $\mathcal{P}_{\mathrm{c}}$ additionally includes a constraint on the autocorrelation metric $\mathcal{T}_{\mathrm{AC}}$, introduced in Eq.\,\eqref{eq:TAC}. 
We choose the value~$1$ as an upper bound, in reference to a perfect reflector, for which $\eta_{\mathrm r}=1$ and $\mathcal{T}_{\mathrm{AC}}=1$.
We solve the two inverse-design problems in Eqs.~\eqref{eq:topopt-unconstrained}
and~\eqref{eq:topopt-constrained}. The optimized designs obtained by topology optimization are shown in Fig.\,\ref{Fig:Statistics}(a) together with their corresponding reflectance spectra, starting from the same random initial density illustrated in Fig.\,\ref{Fig:Setup}. The (normalized) incident spectrum is shown as the red dashed line for reference. Without the constraint, the reflectance spectrum exhibits a prominent narrow spectral feature, similar to a phase-defect resonance known from \(\pi\)-shifted gratings~\cite{piShifted}, and attains a relatively high value of \(\mathcal{T}_{\mathrm{AC}}=8.628\). Enforcing the constraint penalizes the associated long-lag autocorrelation tail and produces a smoother broadband response. This example supports our statement that maximizing Parseval's energy alone is insufficient to capture and remove sharp spectral features, even if the objective reached a (local) optimum.

This first example compared only two optimizations and might thus not be representative of the success of including the autocorrelation metric, as most inverse design problems are known to be non-convex and thus have several local extrema. We therefore performed a statistical evaluation for 100 different initial densities, with and without the constraint on $\mathcal{T}_{\mathrm{AC}}$, summarized in Fig.\,\ref{Fig:Statistics}(b).

From the broadband reflectance statistics in Fig.\,\ref{Fig:Statistics}(b.1), we observe that enforcing the constraint leads to a consistent removal of sharp spectral features in the reflectance spectra compared to the case where the optimization is driven only by maximizing \(\eta_{\mathrm r}\). The distributions of the time-domain metrics in Fig.\,\ref{Fig:Statistics}(b.2) further show that including the constraint does not statistically reduce the optimized reflectance performance \(\eta_{\mathrm r}\): the median values are~\(0.8096\) and~\(0.8093\) for the unconstrained and constrained problems, respectively. This indicates that the constraint has a regularizing effect rather than directly opposing high objective values. 
This is further supported by evaluating the distribution of~\(\mathcal{T}_{\mathrm{AC}}\). For the unconstrained ensemble, \(\mathcal{T}_{\mathrm{AC}}\) is broadly distributed, with a median value of~\(1.854\). However, the best design in terms of~\(\eta_{\mathrm r}\) gives \(\mathcal{T}_{\mathrm{AC}}=1.371\), approaching the prescribed \(\mathcal{T}_{\mathrm{AC}}=1\) bound imposed in the constrained optimizations. Thus, high reflectance is compatible with comparatively low autocorrelation tails, and the constraint mainly guides the optimization toward such designs more consistently.

We further computed the measure of non-discreteness~$M_{\mathrm{nd}}$ from Eq.\,\eqref{eq:M-nd} for both ensembles, which yields a median of~0.04\%~vs~0.30~\% for unconstrained vs constrained optimization. In both cases, the designs tend to converge to a nearly binary grating, and post-binarization of the designs (by thresholding at~$0.5$) did not significantly change their performance:
After thresholding, the maximum paired changes in the unconstrained ensemble were
\(\max|\Delta\eta_{\mathrm r}|=3.58\times10^{-5}\) and
\(\max|\Delta\mathcal{T}_{\mathrm{AC}}|=4.16\times10^{-2}\).
For the constrained ensemble, the corresponding maxima were
\(1.92\times10^{-3}\) and \(3.92\times10^{-2}\), respectively.


\subsection{Autocorrelation metric as a penalization term}
To better understand the interplay between forcing high $\eta_{\mathrm{r}}$ and low $\mathcal{T}_{\mathrm{AC}}$, and its impact on the convergence to a binary grating ($M_{\mathrm{nd}}\xrightarrow{}0$), we now study the optimization problem

\begin{equation}
    \max_{\rho}\;J:= (1-\alpha)\,\eta_{\mathrm r} -
\alpha\, \mathcal{T}_{\mathrm{AC}},
\label{eq:objective-combined}
\end{equation}
using a linear combination of both quantities with a weighting parameter $\alpha\in[0,1]$. For \(\alpha=0\), it reduces to the problem \(\mathcal{P}_{\mathrm{uc}}\) from Eq.\,\eqref{eq:topopt-unconstrained}, whereas for \(\alpha=1\), the objective reduces to minimizing \(\mathcal{T}_{\mathrm{AC}}\) alone. We performed single optimizations for different values of~\(\alpha\), starting from the same random initial density shown in Fig.~\ref{Fig:Setup}. For all values of~\(\alpha\), the objective approached a plateau
within a fixed (optimization-) iteration budget (App.~\ref{sec:appendix}). For \(\alpha\geq0.2\), we note that the density iterates retained a small persistent oscillation rather than settling to a unique fixed design; the neighboring designs nevertheless differed little in objective value and did not affect the overall performance trade-off.

\begin{figure}[!t]
    \centering
    \includegraphics[trim=0mm 2mm 0mm 0mm,width=\columnwidth]{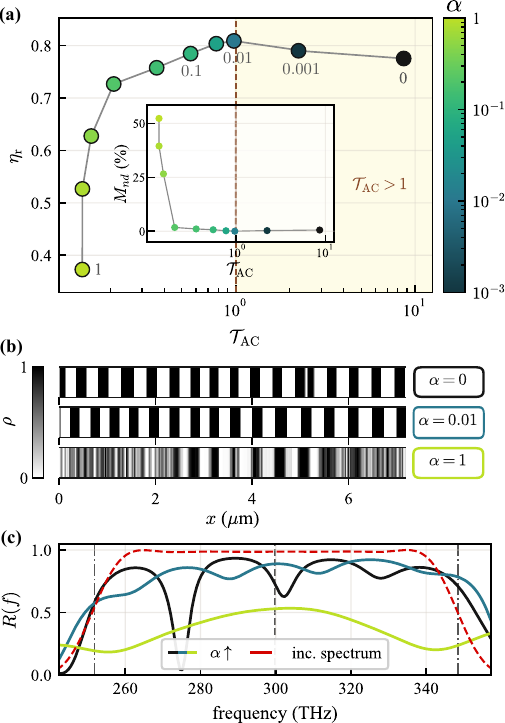}
    \caption{(a)~Pareto plot of $\eta_\mathrm{r}$ vs. $\mathcal{T}_{\mathrm{AC}}$ for different weighting parameters $\alpha$ for the designs evolved under TopOpt problem from~Eq.\,\eqref{eq:objective-combined} starting from the same initial density as in~Fig.\,\ref{Fig:Setup}. The inset plot shows the  $M_{\mathrm{nd}}$ vs. $\mathcal{T}_{\mathrm{AC}}$. (b)~Optimized grating structures for three representative $\alpha$-values, where \mbox{$\alpha=0$} emphasizes the maximization of $\eta_\mathrm{r}$ exclusively, while $\alpha=1$ emphasizes the minimization of $\mathcal{T}_{\mathrm{AC}}$; (c) shows their corresponding reflectance response over the target bandwidth.}
    \label{Fig:Pareto}
\end{figure}

The results are summarized in Fig.\,\ref{Fig:Pareto} in the form of a Pareto plot and three different representative designs shown below together with their respective reflectance spectra. We observe that even a small value of~\(\alpha\) leads to designs with higher~\(\eta_{\mathrm r}\), and that a maximum is reached at \(\alpha=10^{-2}\), for which \(\mathcal{T}_{\mathrm{AC}}\) approaches~\(1\). The designs obtained for \(\alpha=0\) and \(\alpha=10^{-3}\), which
combine lower~\(\eta_{\mathrm r}\) with substantially higher
\(\mathcal{T}_{\mathrm{AC}}\), represent suboptimal local
solutions of the non-convex design problem. Penalizing the partial objective $\mathcal{T}_{\mathrm{AC}}$ even further leads to a decrease of $\eta_{\mathrm r}$ and an increase in $M_{\mathrm{nd}}$, where the designs tend to favor an apodized grating profile, cf. Fig.\,\ref{Fig:BraggGratings}. Apodized Bragg gratings are known to reduce abrupt boundary scattering by smoothly varying the grating strength along the propagation direction, and are commonly used to suppress oscillatory sidelobes in the reflectance spectrum~\cite{Othonos1997Review}.

The Pareto plot together with the evaluation of the spectral performance indicates that a weak penalty on \(\mathcal{T}_{\mathrm{AC}}\) can guide the optimizer toward a better local optimum with higher
\(\eta_{\mathrm r}\), whereas stronger prioritization of spectral
smoothness comes at the cost of high reflectance over the target bandwidth.


\subsection{Discovering a chirped grating by topology optimization}
Finally, we demonstrate the potential for achieving nearly uniform
reflectance with our \(\mathcal{T}_{\mathrm{AC}}\)-penalized
formulation and observe a characteristic spatial pattern in the
optimized grating, thereby confirming what we have shown in one of our previous works: namely, that TopOpt can serve not only as an efficient
design tool, but also as a source of physical insight~\cite{GedeonPowerDissip}.

\begin{figure}[!t]
    \centering
    \includegraphics[trim=0mm 2mm 0mm 0mm,width=\columnwidth]{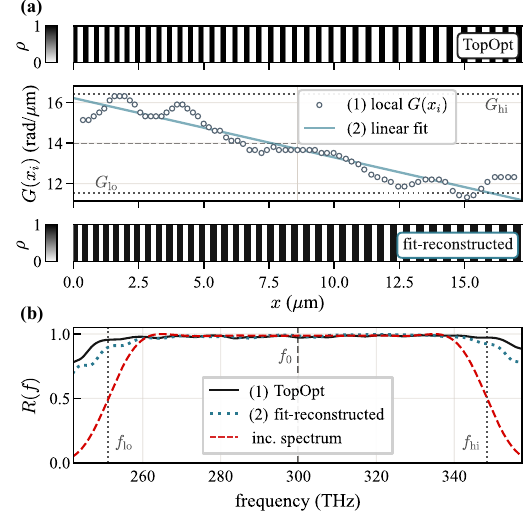}
    \caption{Emergence and parametric reconstruction of a chirped grating from topology optimization. (a)~Top:~thresholded TopOpt design obtained with the weighted objective in Eq.\,\eqref{eq:objective-combined} using \(\alpha=10^{-2}\) and \(L=17.2~\mu\mathrm{m}\). Middle:~local grating vectors \(G(x_i)=2\pi/\Lambda(x_i)\) extracted from the binary design, with the linear fit \(G^{\mathrm{fit}}(x)\) shown as a purple solid line. Bottom:~binary grating reconstructed from the fitted linear chirp using Eq.\,\eqref{eq:parametric-grating}. (b)~Broadband responses of the TopOpt design and the reconstructed grating, showing comparable near-unity reflectance over the target bandwidth.}
    \label{Fig:G-fit}
\end{figure}

We perform a TopOpt run with an extended design length of \(L=17.2~\mu\mathrm{m}\), and use the weighted objective formulation from Eq.\,\eqref{eq:objective-combined} with \(\alpha=10^{-2}\), and starting from a random initial density. The optimized, (post-)binarized structure is shown in Fig.\,\ref{Fig:G-fit}(a), top row. 
To obtain deeper insight into the evolved spatial pattern, we extracted a local period \(\Lambda(x_i)\) at positions \(x_i\) by measuring the distance between neighboring grating edges with the same orientation. The corresponding local grating vector defined as \(G(x_i)=2\pi/\Lambda(x_i)\) is illustrated in Fig.\,\ref{Fig:G-fit}(a) over the entire grating length. Its profile shows an overall decrease from the left to the right design edge and appears approximately antisymmetric with respect to the grating center. We model this dependence by a simple linear function
\begin{equation}
G^{\mathrm{fit}}(x)
=
a_{1}
-
\left(a_{1}-a_{2}\right)\frac{x}{L},
\qquad x\in[0,L],
\label{eq:linear-chirp-fit}
\end{equation}
with positive fitting parameters \(a_{1}\) and \(a_{2}\), minimizing the sum of squared residuals. The values are listed in Table\,\ref{tab:chirp-fit}, together with the source-reference values obtained from the target band \(f_0\pm\Delta f\) by evaluating the first-order Bragg estimate \(G(f)=4\pi n_{\mathrm{eff}}f/c\), with \(n_{\mathrm{eff}}=(\sqrt{\varepsilon_{\mathrm b}}+\sqrt{\varepsilon_{\mathrm d}})/2\). From this, we observe that the fitting values \(a_{1}\) and \(a_{2}\) closely match the Bragg vectors associated with the spectral bounds of the incident pulse. We found that our grating has, in fact, converged to a chirped-like grating, cf.~Fig.\,\ref{Fig:BraggGratings}. Chirped gratings are well known for enabling broadband reflection via a gradual spatial variation in period, where the Bragg condition is satisfied at different spatial positions~\cite{TosiReview}.
\begin{table}[!t]
\caption{Comparison between source-reference grating vectors and the fitted parameters of the linear-chirp reconstruction.}
\label{tab:chirp-fit}
\centering
\renewcommand{\arraystretch}{1.15}
\begin{tabular}{lccc}
\hline
Quantity & Source spectrum & Linear fit & Unit \\
\hline
High edge & \(G_{\mathrm{hi}}=16.425\) & \(a_{1}=16.23\) & \(\mathrm{rad}/\mu\mathrm{m}\) \\
Low edge & \(G_{\mathrm{lo}}=11.532\) & \(a_{2}=11.18\) & \(\mathrm{rad}/\mu\mathrm{m}\) \\
\hline
\end{tabular}
\end{table}

We can use our parametric modeling of the TopOpt design to reconstruct a chirped one. A grating with a spatially varying period can be described through the accumulated local phase~\cite{Accumulation},
\begin{equation}
\phi(x)
=
\phi_0+\int_{0}^{x}G(s)\,ds,
\label{eq:grating-phase}
\end{equation}
where \(\phi_0\) accounts for a spatial shift of the grating pattern. A binary grating with local duty cycle \(D(x)\) can then be obtained by the rule
\begin{equation}
\rho_{\mathrm{bin}}(x)
=
\begin{cases}
1, & \operatorname{mod}\!\left(\phi(x),2\pi\right)<2\pi D(x),\\
0, & \text{otherwise}.
\end{cases}
\label{eq:parametric-grating}
\end{equation}
We reconstruct a linear-chirped grating by inserting \(G^{\mathrm{fit}}(x)\) into Eq.\,\eqref{eq:grating-phase}, with the phase offset \(\phi_0\) chosen by pixel-wise alignment with the TopOpt design. The binary pattern is then obtained from Eq.\,\eqref{eq:parametric-grating} using a fitted duty cycle, which was observed to be approximately constant, \(D(x)\approx D_0=0.46\). The result is shown in the bottom row of Fig.\,\ref{Fig:G-fit}(a). 
The simulated time metrics of the TopOpt design are \(\eta_{\mathrm r}=0.978\) and \(\mathcal{T}_{\mathrm{AC}}=1.0467\), while the reconstructed grating yields slightly lower but comparable values of \(\eta_{\mathrm r}=0.973\) and \(\mathcal{T}_{\mathrm{AC}}=1.043\). The broadband responses shown in Fig.\,\ref{Fig:G-fit}(b) confirm that both designs achieve an almost flat, near-unity reflectance across the entire target bandwidth.

\section{Conclusion}
We addressed a key limitation of broadband time-domain inverse design, where localized spectral features can contribute little to commonly used time-integrated objectives and their spectrally integrated counterparts, by introducing a time-domain metric~\(\mathcal{T}_{\mathrm{AC}}\) that quantifies long-lag autocorrelation energy as a measure of delayed ringing and spectral non-uniformity relative to a broadband reference pulse. Using dielectric Bragg-grating mirrors as an example, we showed that incorporating this metric into a time-domain inverse design scheme can suppress sharp spectral variations in the reflectance spectrum, which may otherwise remain effectively invisible to the optimizer. We found that a weak penalty on the metric does not come at the expense of high reflectance efficiency, and can instead guide the optimizer toward a better local optimum. Strong penalization, however, yields the expected trade-off between spectral uniformity and reflectance efficiency. By enforcing suppression of delayed ringing while targeting broadband reflectance, our optimized structures recovered apodized and chirped designs that are well-known in the field of fiber grating optics~\cite{Othonos1997Review}.

Our approach can be related to the pulse-shaping method introduced in Ref.~\cite{PulseShaping_Lazarov2011}, where the Hilbert envelope of the output signal is optimized to match a prescribed target envelope. In contrast, penalizing $\mathcal{T}_{\mathrm{AC}}$ imposes a weaker constraint: it does not prescribe a specific envelope, leaving the detailed temporal shape, delay, and spreading to the optimizer.

We believe that the metric can serve as an efficient time-domain measure for assessing and suppressing broadband spectral fluctuations and localized spectral features in inverse-designed optical devices. It can be used either as a weighted objective or constraint, or for post-evaluation by quantifying broadband performance from the autocorrelation signature. Beyond the grating optimization used here as a demonstration case, the metric can also be directly incorporated into other, more challenging inverse-design problems, such as wideband near-field coupling~\cite{HassanAntennas}, achromatic focusing~\cite{AdaptivePulse}, or broadband absorption~\cite{GedeonMetasurface}.


\appendices
\section{FDTD and adjoint method}\label{app:TopOpt}\label{sec:appendix}
In the simulations, we solve the one-dimensional Maxwell equations for the field components \(E_y\) and \(H_z\) in time,
\begin{subequations}\label{Eq:Maxwells_1D}
\begin{empheq}[]{align}
\partial_x H_z
+ \varepsilon_{0}\varepsilon_{\mathrm{r}}(\rho)\,\partial_t E_y
&= S_y, \\[1pt]
\mu_0 \,\partial_t H_z+\partial_x E_y
&=0.
\end{empheq}
\end{subequations}
Here, \(S_y\) represents the source term injecting a forward-propagating wave at position~\(x_{\mathrm{inc}}\), and \(\varepsilon_0\) and \(\mu_0\) are the vacuum permittivity and permeability, respectively. The relative permittivity~\(\varepsilon_{\mathrm{r}}(\rho)\) is linked to the density profile~\(\rho(x)\) through the linear interpolation between the background and design media introduced in Sec.~\ref{Sec:Results}; see Fig.\,\ref{Fig:Setup}. This density represents the design field in our topology optimization~(TopOpt). We solve Eq.\,\eqref{Eq:Maxwells_1D} using the finite-difference time-domain~(FDTD) method, in which \(E_y\), \(H_z\) are sampled on a staggered Yee grid and updated using a leapfrog algorithm~\cite{Taflove}. First-order Mur absorbing boundary
conditions are applied at both ends of the computational domain. The density~\(\rho\) and the corresponding relative permittivity~\(\varepsilon_{\mathrm{r}}\) are then discretized accordingly and are collocated with \(E_y\) on the numerical grid. The spatial and temporal step sizes, \(\Delta x\) and \(\Delta t\), are chosen to both satisfy the Courant-Friedrichs-Lewy~(CFL) stability condition and limit numerical dispersion. We choose a sufficiently long simulation time~\(T_{\mathrm{Sim}}=\text{Time steps} \times \Delta t\) for the fields to decay and thereby set the frequency resolution~\(\delta f \propto 1/T_{\mathrm{Sim}}\) of the Fourier-transformed fields presented in our spectral plots.

To compute the gradients of the objective and constraint with respect to the density~$\rho(x)$, we employ the adjoint method in time~\cite{GedeonPowerDissip}. We first solve Eq.\,\eqref{Eq:Maxwells_1D} using the FDTD method and evaluate the objective (and constraint function) from the reflected field recorded at a monitor point~\(x_R<x_{\mathrm{inc}}\) from the excitation of a broadband injected pulse; see Fig.\,\ref{Fig:Setup}. The incident field \(E_{\mathrm{inc}}\) is launched at \(t=0\) using a source signal \(s_0\) that approximates a rectangular frequency spectrum with half-bandwidth \(\Delta f\) through a finite-duration, Hann-windowed sinc pulse modulated by a cosine carrier at the center frequency~\(f_0\),
\begin{equation}
s_0(\tilde{t})=
\begin{cases}
\displaystyle
\operatorname{sinc}(2\Delta f\,\tilde{t})\,
\frac{1}{2}\left[1+\cos\left(\frac{\pi\tilde{t}}{T_{\mathrm p}}\right)\right]
\cos(2\pi f_0\tilde{t}),
& |\tilde{t}|\leq T_{\mathrm p},\\
0, & |\tilde{t}|>T_{\mathrm p},
\end{cases}
\label{eq:source-waveform}
\end{equation}
where \(\tilde{t}=t-t_0\), \(t_0\) is the pulse-center time, and
\(2T_{\mathrm p}=N_{\mathrm{lobes}}/\Delta f\) is the total pulse duration. Increasing \(N_{\mathrm{lobes}}\) increases the pulse duration and sharpens the spectral edges, thereby making the time-domain objective in Eq.\,\eqref{eq:reflection-efficiency} a more accurate approximation of the band-averaged reflectance. However, this comes at the cost of a longer FDTD simulation time. For the signal used in Sec.~\ref{Sec:Results}, with \(f_0=300~\mathrm{THz}\) and \(2\Delta f=105~\mathrm{THz}\), we chose \(N_{\mathrm{lobes}}=6\). For this signal, the common lower integration bound used to compute \(\mathcal{T}_{\mathrm{AC}}\) is \(\tau_{\min}\approx10.14~\mathrm{fs}\). We extracted this value from the incident pulse by applying a Hilbert transform to its autocorrelation function and selecting the lag at which the Hilbert envelope has decayed to \(1\%\) of its peak value, cf.~Sec.~\ref{Sec:TAC}. The autocorrelation is
evaluated by a zero-padded, FFT-based convolution of the time-domain signal with its time-reversed copy using \texttt{NumPy}'s real FFT and inverse real FFT.

After this forward simulation, we solve the corresponding adjoint system backward in time, with the source term~\(S_y\) replaced by the time-reversed derivative of the objective (and constraint function) with respect to the electric field~\(E_y\)~\cite{GedeonPowerDissip}. The sensitivities are obtained from an overlap integral between the forward and adjoint electric fields over time. This gradient information is then passed to a gradient-based optimizer to update~\(\rho(x)\), and the procedure is repeated until the density field and objective value have converged.
\begin{table}[!t]
\caption{FDTD and TopOpt parameters used in the optimizations.}
\label{tab:topopt-fdtd-params}
\centering
\renewcommand{\arraystretch}{1.03}
\setlength{\tabcolsep}{4.5pt}
\begin{tabular}{lccc}
\hline
Parameter & Fig.\,\ref{Fig:Statistics} & Fig.\,\ref{Fig:Pareto} & Fig.\,\ref{Fig:G-fit} \\
\hline
\multicolumn{4}{l}{\textit{FDTD}} \\
\(\Delta x\) (nm) & \(20\) & \(20\) & \(5\) \\
\(\Delta t\) (as) & \(63.38\) & \(63.38\) & \(15.84\) \\
Grid points, full domain & \(701\) & \(701\) & \(4804\) \\
Grid points, design domain & \(360\) & \(360\) & \(3440\) \\
Time steps & \(25\,000\) & \(25\,000\) & \(400\,000\) \\
\hline
\multicolumn{4}{l}{\textit{TopOpt}} \\
Filter radius (cells) & \(2\) & \(2\) & \(4\) \\
Filtered stage iterations & \(160\) & \(160\) & \(250\) \\
Total MMA iterations & \(460\) & \(2000\) & \(650\) \\
\hline
\end{tabular}
\end{table}

As the gradient-based optimizer, we used the Python implementation~\texttt{mmapy} of the method of moving asymptotes (MMA)~\cite{mmapy,SvanbergMMA}. We additionally applied density filtering with a fixed filter radius~\cite{OleFilter}, and used a two-stage continuation strategy consisting of an initial filtered-density stage followed by unfiltered polishing iterations. This strategy was observed to improve convergence, particularly for the constrained optimization problem in Eq.\,\eqref{eq:topopt-constrained}. Details on the chosen MMA parameters and convergence histories are provided in our GitHub repository and dataset~\cite{Grating_TimeOpt_Github, dataset}.

Table~\ref{tab:topopt-fdtd-params} summarizes the FDTD and TopOpt configuration settings used for our simulations and optimizations presented in Sec.~\ref{Sec:Results}.

\bibliographystyle{IEEEtran}
\bibliography{main}

\end{document}